\documentclass[oldversion]{aa}  
\usepackage{graphicx}
\usepackage{txfonts}

\def\etal{{et\,al.}\ }
\def\kpd{KPD\,0005+5106}
\def\elf{PG\,1159$-$035}
\def\ev{PG\,1144$+$005}
\def\vier{PG\,1424$+$535}
\def\fuenf{PG\,1520$+$525}
\def\re{RE\,J0503$-$289}

\newcommand{\Teff}{$T\mathrm{\hspace*{-0.4ex}_{eff}}$\,}
\newcommand{\logg}{$\log\,g$\hspace*{0.5ex}}

\begin{document}
\title{Iron abundance in the prototype PG\,1159 star, GW Vir pulsator
  PG\,1159$-$035, and related objects\thanks{Based on observations made with the
  NASA-CNES-CSA Far Ultraviolet Spectroscopic Explorer. FUSE was operated for
  NASA by the Johns Hopkins University under NASA contract NAS5-32985. Based on
  observations with the NASA/ESA Hubble Space Telescope, obtained at the Space
  Telescope Science Institute, which is operated by the Association of
  Universities for Research in Astronomy, Inc., under NASA contract
  NAS5-26666. }  }

\author{K\@. Werner\inst{1}
   \and T\@. Rauch\inst{1}
   \and J.W\@. Kruk\inst{2}
   \and R.L\@. Kurucz\inst{3}}

\institute{Institute for Astronomy and Astrophysics, Kepler Center for Astro and
Particle Physics,  Eberhard Karls Universit\"at T\"ubingen, Sand~1, 72076
T\"ubingen, Germany, \email{werner@astro.uni-tuebingen.de} \and NASA Goddard
Space Flight Center, Greenbelt, MD 20771, USA \and  Harvard-Smithsonian Center
for Astrophysics, 60 Garden Street, Cambridge, MA 02138, USA  }

\date{Received 31 March 2011 / Accepted 13 May 2011}

\authorrunning{K. Werner \etal}
\titlerunning{Iron abundance in PG\,1159$-$035 and
  related objects}

\abstract{We performed an iron abundance determination of the hot,
  hydrogen deficient post-AGB star \elf, which is the prototype of the PG\,1159 spectral
  class and the GW~Vir pulsators, and of two related objects (\fuenf, \ev), based
  on the first detection of \ion{Fe}{viii} lines in stellar photospheres. In
  another PG\,1159 star, \vier, we detect \ion{Fe}{vii} lines. In all four
  stars, each within \Teff = 110\,000 -- 150\,000\,K, we find a solar iron
  abundance. This result agrees with our recent abundance analysis of
  the hottest PG\,1159 stars (\Teff = 150\,000 -- 200\,000\,K) that exhibit
  \ion{Fe}{x} lines. On the whole, we find that the PG\,1159 stars are not
  significantly iron deficient, in contrast to previous notions.  }

\keywords{Stars: individual: \elf\ --
          Stars: abundances -- 
          Stars: atmospheres -- 
          Stars: evolution  -- 
          Stars: AGB and post-AGB --
          White dwarfs}

\maketitle
%
%________________________________________________________________

\begin{figure*}
  \centering 
  \includegraphics[width=0.9\textwidth]{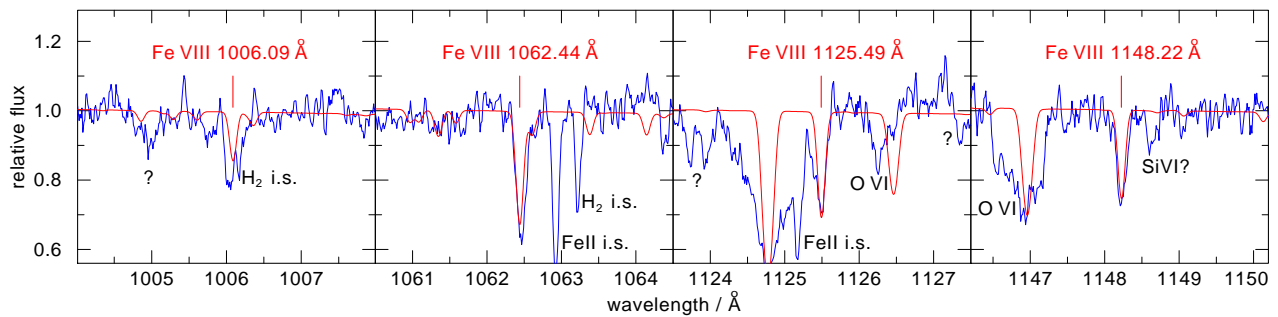}
  \caption{\ion{Fe}{viii} lines detected in \elf. Overplotted is the final model
  with solar iron abundance (model parameters: \Teff = 140\,000\,K, \logg = 7,
  He/C/O/Ne = 0.32/0.48/0.17/0.02; mass fractions).
}\label{fig:fe8_pg1159}
\end{figure*}

\begin{figure*}
  \centering  \includegraphics[width=0.6\textwidth]{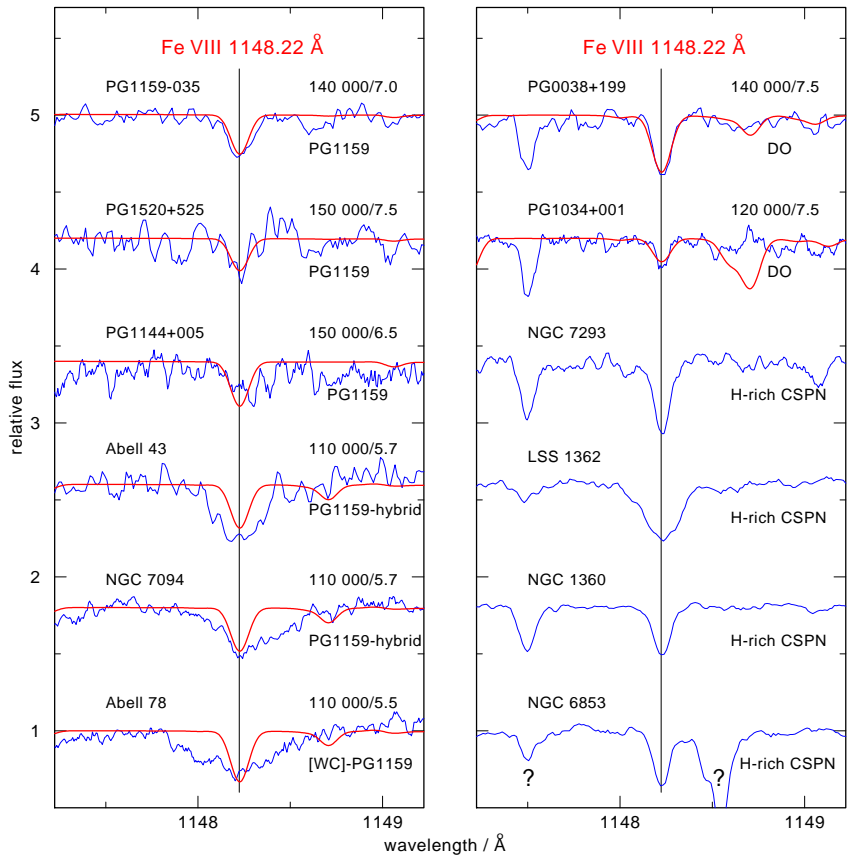}
  \caption{The \ion{Fe}{viii} $\lambda$\,1148.22\,\AA\ line in PG\,1159 stars and
  other objects. Left panel, from top: three PG\,1159 stars, two
  PG\,1159-hybrids, and a [WC]--PG\,1159 transition object. Right panel: two hot
  DOs and four H-rich central stars. Depicted numbers denote temperature and
  gravity of the overplotted (red) models. They have solar iron abundance except
  for the two DOs where Fe is ten times solar.  }\label{fig:fe8_allstars}
\end{figure*}

\section{Introduction}
\label{intro}

PG\,1159 stars are hot hydrogen deficient post-AGB stars (Werner \& Herwig
2006). In the Hertzsprung-Russell diagram, they cover a region comprising the
hottest central stars of planetary nebulae and white dwarfs (\Teff = 75\,000 --
200\,000\,K, \logg = 5.5 -- 8). Their H deficiency is most probably the result
of a late He-shell flash. Their envelopes are mainly composed of He, C, and O,
with rather diverse abundance patterns (He = 0.30 -- 0.85, C = 0.13 -- 0.60, O =
0.02 -- 0.20, mass fractions).

The prototype \elf\ (= GW~Vir) was discovered in the Palomar Green survey
(Wesemael \etal 1985). Subsequently it was found that the star is variable
(McGraw \etal 1979) and it also became the prototype of the GW~Vir stars, which
are non-radial multimode g-mode pulsators. Besides the Sun, \elf\ is probably
the star that is best studied with asteroseismic methods (Costa \etal 2008). One
of the key questions related to these pulsators concerns the driving mechanism,
because the instability strip occupied by them is not ``pure'' like the ZZ~Ceti
strip, meaning that it also contains non-variable PG\,1159 stars.

The primary pulsation driver is cyclic ionisation of C and O (Starrfield \etal
1984). The location of the instability strip is ``fuzzy'', because the red and blue
edges of the strip depend on the He/C/O abundance ratio in the driving region;
too high an He abundance poisons pulsations (Quirion \etal 2007). Another species
that supports pulsation driving is iron, therefore, a subsolar iron abundance would
narrow the instability strip.

\elf\ (\Teff = 140\,000\,K, \logg = 7), together with a near spectroscopic twin,
the non-pulsator \fuenf\ (\Teff = 150\,000\,K, \logg = 7.5), potentially defines
the blue edge of the GW~Vir strip, provided their envelope chemical composition
is similar. Considerable effort was put into spectroscopic analyses to derive
temperature, gravity, and composition of these twin stars. One remaining open
question is the iron abundance. For some PG\,1159 stars, including the twins,
there were claims of iron deficiency (Jahn \etal 2007).

Spectroscopically, the iron abundance in PG\,1159 stars is difficult to assess.
Hitherto, the main tool were ultraviolet \ion{Fe}{vii} lines, well known from
observations of hot hydrogen-rich central stars of planetary nebulae. Because of
the high effective temperatures, these lines are predicted to be rather weak or
even undetectable in PG\,1159 stars. The previously mentioned Fe deficiency was
based on not detecting \ion{Fe}{vii} lines. UV lines from higher
ionisation stages of iron were unknown until recently, when \ion{Fe}{x} lines
were detected in five of the very hottest (\Teff $\geq$ 150\,000\,K) PG\,1159
stars (Werner \etal 2010). A solar iron abundance was derived.

In this paper, we announce the detection of \ion{Fe}{viii} lines in FUSE spectra
of three medium-hot (\Teff = 140\,000 -- 150\,000\,K) PG\,1159 stars, including
the prototype and its twin. They serve as a tool for determining the iron
abundance, closing the gap between the coolest  (\Teff $\la$ 140\,000\,K)
PG\,1159 stars, where we should be able to detect \ion{Fe}{vii}, and the very
hottest objects exhibiting \ion{Fe}{x}. For the first time, we present an iron
abundance determination of \elf.  We carefully re-assess archival FUSE and HST
spectra of \elf\ to look for weak, previously undetected \ion{Fe}{vii}
lines. This search is extended to the cooler object \vier\ (\Teff = 110\,000\,K,
\logg = 7), where the non-detection of these lines would mean a Fe deficiency of
one dex (Reiff \etal 2008).  We also report on \ion{Fe}{viii} lines in other hot
H deficient and H rich post-AGB stars.

In the following section, we present the detection of \ion{Fe}{viii} lines in
\elf\ and other objects (Sect.\,\ref{observations}). Then we describe our model
atmospheres (Sect.\,\ref{modeling}) and the spectroscopic iron abundance
analysis of four PG\,1159 stars (Sect.\,\ref{analysis}), and we conclude in
Sect.\,\ref{conclusions}.

\begin{figure*}[ht]
  \centering  \includegraphics[width=0.9\textwidth]{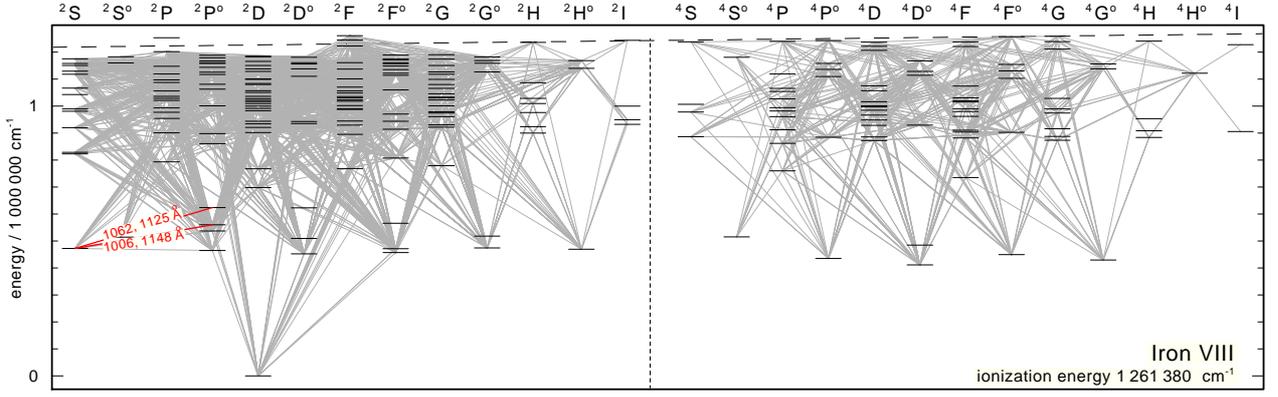}
  \caption{Grotrian diagram of \ion{Fe}{viii}. 
For clarity, it is drawn from Opacity Project (OP) data that represent a small
subset of the Kurucz dataset utilised in our computations.  The OP level
energies differ from Kurucz values.  The transitions giving rise to the observed
lines are indicated.  }\label{fig_modelatom}
\end{figure*}

\begin{figure*}[ht]
  \centering  \includegraphics[width=0.7\textwidth]{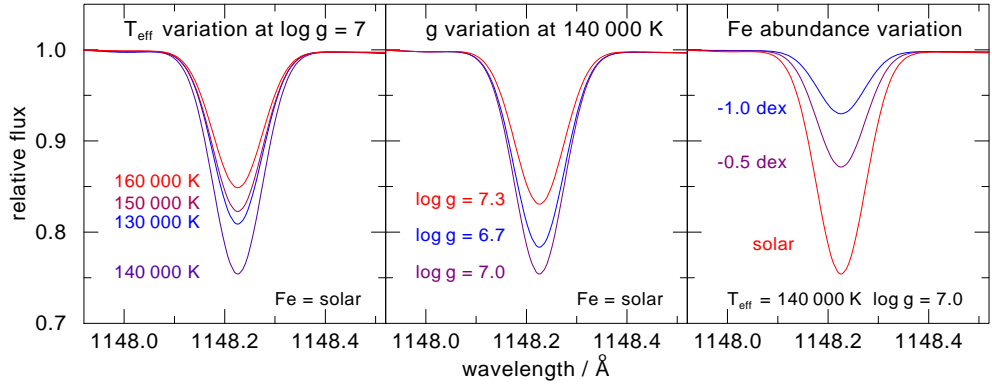}
  \caption{Effects of model parameter variations on the \ion{Fe}{viii}
  $\lambda$\,1148.22\,\AA\ line profile. The line strength is maximum at the
  parameters of \elf\ (\Teff = 140\,000\,K, \logg = 7); other abundances are
  He/C/O/Ne = 0.32/0.48/0.17/0.02. }\label{fig_variation}
\end{figure*}

\begin{figure*}
  \centering 
  \includegraphics[width=0.8\textwidth]{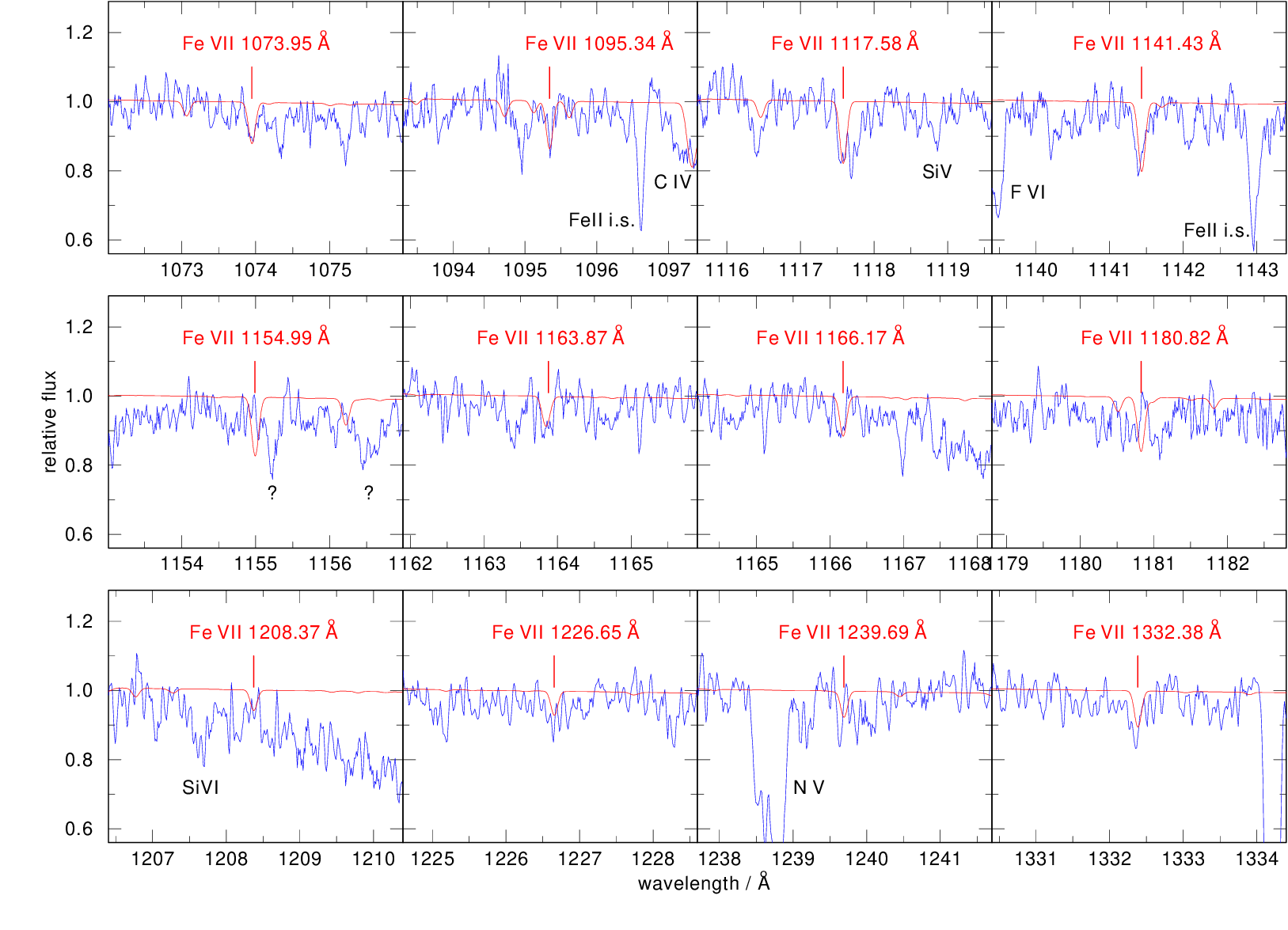}
  \caption{\ion{Fe}{vii} lines in \elf. Overplotted is the final model
  with solar iron abundance (model parameters like in
  Fig.\,\ref{fig:fe8_pg1159}). FUSE data are used for $\lambda < 1200$\,\AA\ and HST
  data otherwise.
}\label{fig:fe7_pg1159}
\end{figure*}

\begin{figure*}[ht]
  \centering 
  \includegraphics[width=0.8\textwidth]{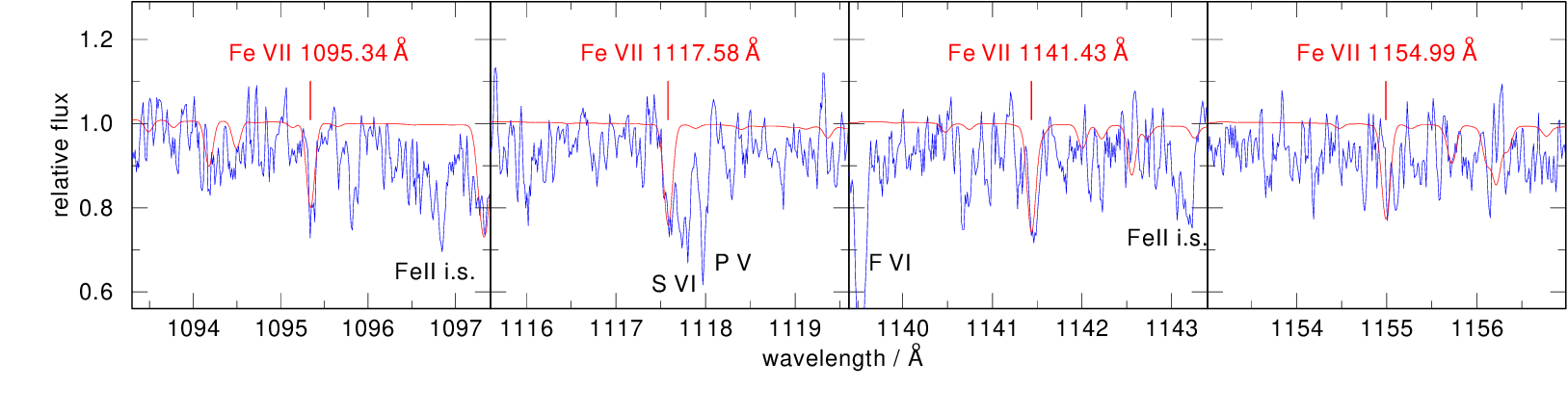}
  \caption{\ion{Fe}{vii} lines in \vier. Overplotted is a model with solar iron
  abundance (model parameters: \Teff = 110\,000\,K, \logg = 7,
  He/C/O/Ne = 0.49/0.43/0.06/0.02). }\label{fig_pgvier}
\end{figure*}

\section{Observations and line identifications}
\label{observations}

Recently, Landi \& Young (2010) have been able to identify four \ion{Fe}{viii}
coronal emission lines in the $\lambda$~1000 -- 1200\,\AA\ region of the quiet
Sun, in spectra obtained with the SOHO/SUMER instrument. This prompted us to
look for accordingly photospheric lines in FUSE spectra of PG\,1159 stars. All
four \ion{Fe}{viii} lines are present in the prototype \elf\
(Fig.\,\ref{fig:fe8_pg1159}).  We find \ion{Fe}{viii} lines in two more PG\,1159
stars and in several other hot (pre-) white dwarfs as well. From these objects
we display the region around \ion{Fe}{viii} $\lambda$\,1148.22\,\AA\ in
Fig.\,\ref{fig:fe8_allstars}.

The two other PG\,1159 stars have slightly higher temperatures than the
prototype (\fuenf: \Teff = 150\,000\,K, \logg = 7.5; \ev: \Teff = 150\,000\,K,
\logg = 6.5). Different PG\,1159 stars from which FUSE spectra exist are
obviously too hot or too cool to exhibit \ion{Fe}{viii} lines. In particular,
these are the hottest ones exhibiting \ion{Fe}{x} lines mentioned in the
introduction (\Teff $\geq$ 150\,000\,K), and the cooler object \vier\ (\Teff =
110\,000\,K) that will be discussed below (Sect.\,\ref{sect42}).

The central stars Abell~43 and NGC\,7094 are hybrid-PG\,1159 stars
(i.e. exhibiting H-Balmer lines), and Abell~78 is a [WC]--PG\,1159 transition
object. They all have low surface gravity, and the extraordinary wide profiles
indicate that the \ion{Fe}{viii} lines are strongly affected by a stellar
wind. The low surface gravity also favours the appearance of \ion{Fe}{viii}
although \Teff\ of these stars is relatively low ($\approx$ 110\,000\,K).

Two of the hottest known DO white dwarfs, PG0038+199 with \Teff =
115\,000\,K and PG1034+001 with 100\,000\,K (Dreizler \& Werner 1996), display
\ion{Fe}{viii} lines. Comparison with preliminary model calculations indicates a
necessary upward revision of the temperatures by $\approx$\,20\,000\,K in order
to reproduce these lines.  We detect \ion{Fe}{viii} in neither \kpd, the
hottest DO (200\,000\,K; Wassermann \etal 2010), nor in cooler DOs like \re\
(70\,000\,K, Dreizler \& Werner 1996).

We also find \ion{Fe}{viii} lines in several hydrogen-rich central stars. Four
very prominent examples are displayed in Fig.\,\ref{fig:fe8_allstars}: NGC\,7293
(\Teff = 120\,000\,K, \logg = 6.3), LSS\,1362 (\Teff = 114\,000\,K, \logg =
5.7), NGC\,1360 (\Teff = 97\,000\,K, \logg = 5.3), NGC\,6853 (\Teff =
126\,000\,K, \logg = 6.5); the parameters are from Hoffmann \etal (2005).

\begin{table}
\begin{center}
\caption{Wavelengths and oscillator strengths $f_{ij}$ of \ion{Fe}{viii} lines.
\label{tab:levels}}
\begin{tabular}{l l l l l }
      \hline
      \hline
      \noalign{\smallskip}
 Line &   $\lambda_{\rm Kurucz}$/\AA &  $\lambda_{\rm Landi}$/\AA & $f_{ij}$ \\
      \noalign{\smallskip}
      \hline
      \noalign{\smallskip}
$4{\rm s}\ ^2{\rm S}_{1/2}-3{\rm d}^2\ ^2{\rm P}^{\rm o}_{1/2}$ & 1006.087 & 1006.015 & 0.0169 \\    
$4{\rm s}\ ^2{\rm S}_{1/2}-4{\rm p}\ \ ^2{\rm P}^{\rm o}_{3/2}$ & 1062.440 & 1062.463$^{\rm (1)}$& 0.401  \\
$4{\rm s}\ ^2{\rm S}_{1/2}-4{\rm p}\ \ ^2{\rm P}^{\rm o}_{1/2}$ & 1125.492 & 1125.546   & 0.216   \\   
$4{\rm s}\ ^2{\rm S}_{1/2}-3{\rm d}^2\ ^2{\rm P}^{\rm o}_{3/2}$ & 1148.224 & 1148.223 & 0.0828  \\  
      \noalign{\smallskip}
      \hline
     \end{tabular}
\\(1) mean value from two measurements
\end{center}
\end{table}

\section{Model atmospheres and synthetic spectra}
\label{modeling}

For our analysis we use a grid of line-blanketed non-LTE model atmospheres,
which is described in detail in Werner \etal (2004). In essence, the models
include the main photospheric constituents, namely He, C, O, Ne, and
occasionally H. NLTE line formation iterations for the iron population densities
were computed on these model structures, i.e., keeping fixed temperature and
density structure. For details on the used iron model atom, see Wassermann \etal
(2010). We employ new versions of iron datasets (Kurucz
2009)\footnote{http://kurucz.harvard.edu/atoms.html}. They include many more
levels and lines, in particular the four \ion{Fe}{viii} lines discussed in this
paper.

Properties of the newly detected \ion{Fe}{viii} lines are listed in
Table~\ref{tab:levels}. They all arise from the same lower level. We specify the
Kurucz wavelengths, as well as those measured by Landi \& Young (2010). The
differences are all smaller than 0.1\,\AA. The largest deviation (0.072\,\AA) is
shown by the 1006\,\AA\ line. The Kurucz wavelengths should be more accurate
than the measured wavelengths 
since the energy levels involved were determined from more than one line. 
We also list the f-values from the
Kurucz data. A simplified Grotrian diagram indicating the observed line
transitions is shown in Fig.\,\ref{fig_modelatom}.

We computed a small model grid in order to study the dependence of the
\ion{Fe}{viii} lines on \Teff, \logg, and Fe abundance. The result for
$\lambda$\,1148\,\AA\ is displayed in Fig.\,\ref{fig_variation}, and the other lines
behave similarly. It turns out that effective temperature and gravity of \elf\ are
the most favourable for the detection of \ion{Fe}{viii}. It also explains why
\ion{Fe}{viii} lines are not seen in objects that are much cooler or hotter.

\begin{table}
\begin{center}
\caption{New \ion{Fe}{vii}, \ion{Fe}{viii}, and \ion{Ne}{vi} lines detected in
   \elf.}
\label{tab:new-lines} 
\begin{tabular}{llc}
\hline 
\hline 
\noalign{\smallskip}
Wavelength / \AA & Ion & Transition \\ \hline
\noalign{\smallskip} 
1006.09 & \ion{Fe}{viii}& $4{\rm s}\ ^2{\rm S}_{1/2}-3{\rm d}^2\ ^2{\rm P}^{\rm o}_{1/2}$ \\
1062.44 & \ion{Fe}{viii}& $4{\rm s}\ ^2{\rm S}_{1/2}-4{\rm p}\ \ ^2{\rm P}^{\rm o}_{3/2}$\\
1073.95 & \ion{Fe}{vii} & $4{\rm s}\ ^1{\rm D}-4{\rm p}\ ^1{\rm P}^{\rm o}      $ \\
1095.34 & \ion{Fe}{vii} & $4{\rm s}\ ^3{\rm D}_3-4{\rm p}\ ^3{\rm P}^{\rm o}_2  $ \\ 
1117.58 & \ion{Fe}{vii} & $4{\rm s}\ ^1{\rm D}-4{\rm p}\ ^1{\rm F}^{\rm o}      $ \\ 
1125.49 & \ion{Fe}{viii}& $4{\rm s}\ ^2{\rm S}_{1/2}-4{\rm p}\ \ ^2{\rm P}^{\rm o}_{1/2}$  \\
1141.43 & \ion{Fe}{vii} & $4{\rm s}\ ^3{\rm D}_3-4{\rm p}\ ^3{\rm F}^{\rm o}_4  $ \\
1148.22 & \ion{Fe}{viii}& $4{\rm s}\ ^2{\rm S}_{1/2}-3{\rm d}^2\ ^2{\rm P}^{\rm o}_{3/2}$ \\
1154.99 & \ion{Fe}{vii} & $4{\rm s}\ ^3{\rm D}_2-4{\rm p}\ ^3{\rm F}^{\rm o}_3  $ \\
1163.88 & \ion{Fe}{vii} & $4{\rm s}\ ^3{\rm D}_2-4{\rm p}\ ^3{\rm D}^{\rm o}_3  $ \\
1166.17 & \ion{Fe}{vii} & $4{\rm s}\ ^3{\rm D}_1-4{\rm p}\ ^3{\rm F}^{\rm o}_2  $ \\
1180.82 & \ion{Fe}{vii} & $4{\rm s}\ ^3{\rm D}_3-4{\rm p}\ ^3{\rm D}^{\rm o}_3  $ \\
1226.65 & \ion{Fe}{vii} & $4{\rm s}\ ^3{\rm D}_3-4{\rm p}\ ^3{\rm D}^{\rm o}_2  $ \\
1239.69 & \ion{Fe}{vii} & $4{\rm s}\ ^3{\rm D}_1-4{\rm p}\ ^3{\rm D}^{\rm o}_1  $ \\
1332.38 & \ion{Fe}{vii} & $4{\rm s}\ ^1{\rm D}-4{\rm p}\ ^1{\rm D}^{\rm o}      $ \\ 
1645.06 & \ion{Ne}{vi}  & $3{\rm s}\ ^4{\rm P}^{\rm o}_{1/2}-3{\rm p}\ ^4{\rm P}_{3/2}$ \\
1645.59 & \ion{Ne}{vi}  & $3{\rm s}\ ^4{\rm P}^{\rm o}_{3/2}-3{\rm p}\ ^4{\rm P}_{5/2}$ \\
1654.01 & \ion{Ne}{vi}  & $3{\rm s}\ ^4{\rm P}^{\rm o}_{1/2}-3{\rm p}\ ^4{\rm P}_{1/2}$ \\
1657.16 & \ion{Ne}{vi}  & $3{\rm s}\ ^4{\rm P}^{\rm o}_{3/2}-3{\rm p}\ ^4{\rm P}_{3/2}$ \\
1666.24 & \ion{Ne}{vi}  & $3{\rm s}\ ^4{\rm P}^{\rm o}_{3/2}-3{\rm p}\ ^4{\rm P}_{1/2}$ \\
1667.82 & \ion{Ne}{vi}  & $3{\rm s}\ ^4{\rm P}^{\rm o}_{5/2}-3{\rm p}\ ^4{\rm P}_{5/2}$ \\
1679.67 & \ion{Ne}{vi}  & $3{\rm s}\ ^4{\rm P}^{\rm o}_{5/2}-3{\rm p}\ ^4{\rm P}_{3/2}$ \\
\noalign{\smallskip} \hline
\end{tabular} 
\end{center}
\vspace{-3mm}
This table augments the UV line list of Jahn \etal (2007), their Table~2.
\end{table}

\section{Iron abundance analysis}
\label{analysis}

\subsection{\elf}

Figure~\ref{fig:fe8_pg1159} shows \ion{Fe}{viii} lines profiles computed from a
solar Fe abundance model for \elf\ compared to the observation. The fit is
satisfactory, and a comparison with the Fe variation shown in the right panel of
Fig.\,\ref{fig_variation} clearly rules out a significant iron deficiency.

In contrast, we previously decided there is an iron deficiency of $>$\,$0.7$~dex from
not detecting \ion{Fe}{vii} lines (Jahn \etal 2007), so we need to
address this question again here. A close inspection of the FUSE and HST spectra
reveals a number of weak \ion{Fe}{vii} lines (Fig.\,\ref{fig:fe7_pg1159}), which
are fitted by our solar Fe abundance model.

The reason we rejected the \ion{Fe}{vii} detection in our previous work was
the apparent absence of the two strong predicted lines at
$\lambda\lambda$\,1154.99 and 1180.82\,\AA\ in the FUSE data. The cause of the
non-detection remains unclear. In particular, we carefully re-addressed the
wavelength calibration. We are confident that the accuracy of the wavelengths is
at least 0.02\,\AA. We also think that the oscillator strengths of these lines
are correct because, together with other \ion{Fe}{vii} lines, they are rather
prominent in spectra of H-rich central stars of planetary nebulae (e.g. Rauch
\etal 2007). Either way, the simultaneous fit of the detected \ion{Fe}{vii} and
\ion{Fe}{viii} lines independently confirms the validity of \Teff\ and \logg\
derived in earlier work.

In Table~\ref{tab:new-lines} we list the iron lines newly detected in \elf,
together with lines from an \ion{Ne}{vi} multiplet
(NIST\footnote{http://physics.nist.gov/pml/data/asd.cfm} wavelengths) that we
have discovered in the HST/STIS spectrum during the present
analysis. This table complements the UV line list presented by Jahn \etal (2007).

In Table~\ref{tab:results} we summarise the photospheric parameters of \elf\
from spectroscopic analyses and derived quantities. In comparison to Jahn \etal
(2007), the table is improved for both the Fe abundance and the
re-determination of mass, luminosity, and distance based on more realistic
evolutionary tracks of Miller Bertolami \& Althaus (2006).

\begin{table}
\begin{center}
\caption{Parameters of \elf. }
\label{tab:results} 
\begin{tabular}{cccc} \hline \hline 
\noalign{\smallskip}
Parameter   & Result                 & Abundances    & Ref.\\
            &                        & (solar units) &     \\
\noalign{\smallskip}
\hline
\noalign{\smallskip}
\Teff / K   & 140\,000 $\pm$ 5000     &                       & (1), (2)\\
\logg / cm s$^{-2}$ & 7.0  $\pm$ 0.5          &                       & (1), (2) \\
\noalign{\smallskip}
H           & $\le 0.02$             & $\le 0.027$           & (2) \\
He          & 0.33                   & 1.3                   & (2) \\
C           & 0.48                   & 203                   & (2) \\
N           & 0.001                  & 1.4                   & (2) \\
O           & 0.17                   & 30                    & (2) \\
F           & $3.2  \cdot 10^{-6}$   & 6.3                   & (2) \\
Ne          & 0.02                   & 16                    & (2) \\
Si          & $3.6 \cdot  10^{-4}$   & 0.54                  & (2)  \\
P           & $\le 6.4 \cdot 10^{-6}$& $\le 1.1$             & (2)  \\
S           & $5.0 \cdot  10^{-6}$   & 0.016                 & (2)  \\
Fe          & $1.3 \cdot 10^{-3}$    & 1.0                   & (1) \\
%\noalign{\smallskip}
%$n_{\rm H}$ / cm$^{-2}$&$1.5 \cdot 10^{20}$&                 & (2) \\
%E(B-V)      & 0.0                    &                       & (2) \\
\noalign{\smallskip}
$M/M_\odot$& $0.536^{+0.068}_{-0.010}$&                        & (1) \\
\noalign{\smallskip}
log $L/L_\odot$& $2.58^{+0.29}_{-0.29}$&                       & (1) \\
\noalign{\smallskip}
$d$/kpc    & $0.750^{+0.334}_{-0.585}$  &                        & (1) \\
\noalign{\smallskip}
\hline
\end{tabular} 
\end{center}
\vspace{-3mm}
Element abundances
  are given in mass fractions (2nd column) and relative to solar abundances
  (Asplund \etal 2009 values; 3rd column). References: 
(1) this work,
(2) Jahn \etal (2007) and references therein, 
(3) Miller Bertolami \& Althaus (2006).
\end{table}

\subsection{\ev, \fuenf, \vier}\label{sect42}

The two other PG\,1159 stars in which we found \ion{Fe}{viii} lines are \ev\
(\Teff = 150\,000\,K, \logg = 6.5) and \fuenf\ (\Teff = 150\,000\,K, \logg =
7.5).  The lines are fitted with profiles from models with solar iron abundance
(Fig.\,\ref{fig:fe8_allstars}). We do not detect \ion{Fe}{vii} lines in these
stars, because they are hotter than \elf\ so that we expect weaker lines, and
because the S/N of the available FUSE spectra is worse. The abundances of
the models shown in Fig.\,\ref{fig:fe8_allstars} are He/C/O/Ne =
0.43/0.38/0.17/0.02 for \fuenf\ and 0.38/0.58/0.02/0.02 for \ev.

The fourth PG\,1159 star considered in the present study, \vier, is too cool to
exhibit \ion{Fe}{viii} lines. The star is interesting because an iron deficiency
of at least 1 dex was concluded from the claimed absence of \ion{Fe}{vii} (Reiff
\etal 2008). Similar to the case of \elf, we again addressed this question and
found that \ion{Fe}{vii} lines are present after all. In Fig.\,\ref{fig_pgvier}
we show a selection of \ion{Fe}{vii} lines from the FUSE spectrum compared to a
solar iron abundance model. The match is very good.

\section{Summary and conclusions}
\label{conclusions}      

Our analysis of \ion{Fe}{vii} and \ion{Fe}{viii} lines in the FUSE spectra of four
PG\,1159 stars results in solar iron abundances. Recent work on five hotter
PG\,1159 stars exhibiting \ion{Fe}{x} lines (Werner \etal 2010) arrived at the
same result. This set of nine stars comprises four objects, which previously
were supposedly iron deficient: \elf, \fuenf, \vier, K\,1-16 (Miksa \etal
2002, Jahn \etal 2007). The reason for this contrary result is twofold:
an underestimation of \Teff\ in the case of K\,1-16 (Werner \etal 2010) and
problems with the identification of the inherently weak lines from \ion{Fe}{vii}
as described in the present work.  \ion{Fe}{vii}  was the only relevant
ionisation stage with accurately known line positions at that time when earlier
analyses were performed.

There are still two objects left with seemingly strong iron deficiency. These
are the hybrid-PG\,1159 star NGC~7094 and the [WC]--PG\,1159 transition object
Abell~78 (Miksa \etal 2002). In both stars we have discovered strong
\ion{Fe}{viii} lines (Fig.\,\ref{fig:fe8_pg1159}). They are much broader than
predicted from our static models. The reason is most probably that the lines
form in the stellar wind of these low-gravity (i.e. high-luminosity) central
stars. The same mechanism could hamper the detection of weak \ion{Fe}{vii}
lines, whose apparent absence the assertion of Fe deficiency was based
on. Because of the prominent \ion{Fe}{viii} lines, we may speculate that the
iron abundance in these objects is about solar, too, but a detailed analysis
with expanding model atmospheres is required.  Our results ease the problem of
explaining the previously believed extreme iron deficiency with stellar evolution
models, which do not predict such large Fe depletions by neutron captures in the
intershell region of AGB stars.

Two of the objects investigated in this study (\elf\ and \fuenf) have rather
similar parameters and they can be regarded as a fixed point for the blue edge
of the GW~Vir instability strip, at least for a particular chemical envelope
composition. Within error limits, they have the same atmospheric abundance
pattern (in particular the Fe abundance) and, thus, differences in the pulsation
driving behaviour should only result from differences in \Teff\ and \logg.

Our correct match of the \ion{Fe}{vii}/\ion{Fe}{viii} ionisation balance
corroborates the previously determined parameters for the pulsator \elf: \Teff =
140\,000\,K, \logg = 7.0. The non-pulsator \fuenf\ has \Teff = 150\,000\,K,
\logg = 7.5. These parameters are confirmed by an analysis of its Chandra X-ray
spectrum (Adamczak et al., in prep.).

\begin{acknowledgements} 
T.R. is supported by the German Aerospace Centre (DLR) under grant 05\,OR\,0806.
Some of the data presented in this paper were obtained from the Multimission
Archive at the Space Telescope Science Institute (MAST). STScI is operated by
the Association of Universities for Research in Astronomy, Inc., under NASA
contract NAS5-26555. Support for MAST for non-HST data is provided by the NASA
Office of Space Science via grant NNX09AF08G and by other grants and contracts.
\end{acknowledgements}

\end{document}